\documentstyle[aps,twocolumn,prl]{revtex}

\begin{document}
\draft
\title{A Consistent Picture for Resonance Neutron Peak and ARPES Spectra in High-$
T_{c}$ Superconductors }
\author{Jian-Xin Li$^{1}$, Chung-Yu Mou$^{2}$, T.K.Lee$^{3,4}$}
\address{
1. National Laboratory of Solid States of Microstructure and Department of\\
Physics, Nanjing University, Nanjing 210093, China}
\address{
2. Department of Physics, National Tsing Hua University, Hsinchu 30043,\\
Taiwan}
\address{
3. Institute of Physics, Academia Sinica, Taipei 11529, Taiwan}
\address{
4. National Center for Theoretical Sciences, P.O.Box 2-131, Hsinchu, Taiwan}
\maketitle

\begin{abstract}
The spectra observed in ARPES measurements are examined together with
the resonance peak observed in neutron scatterings, based on the slave-boson
approach to $t-t^{\prime }-J$ model. We show that the peak/dip/hump
features arise from the scattering of electrons by collective spin
excitations which, at the same time, give rise to the neutron resonance
mode. The doping dependences and the dispersions of the peak/dip/hump
positions are shown to be consistent with experiments. Our results indicate
that recently observed $\cos (6\theta )$ deviation from the pure $d$-wave
also result from the renormalization by spin fluctuations.
\end{abstract}


\preprint{} %\widetext 

Both angle resolved photoemission spectroscopy (ARPES) and neutron
scattering experiments have played important roles in the studies of high-$
T_{c}$ superconductors. It has been shown by ARPES that the spectral
lineshape possesses a peak/dip/hump structure in the superconducting(SC)
state~\cite{des,ran}. The anomalous momentum, temperature and doping
dependences of the spectral lineshape suggest that electrons are strongly
coupled to collective excitations centered at $(\pi ,\pi )$ and these
collective excitations are related to the pairing interaction~\cite{shen}.
On the other hand, the most prominent feature of the spin susceptibility
observed in neutron scattering studies is the sharp resonance peak at $(\pi
,\pi )$ in the superconducting or pseudogap states~\cite{ros,moo,fong}. It
has been speculated~\cite{shen,nor1}, from a comparison of these two kinds
of experimental data, that the collective excitations are the resonance
modes in the neutron scattering experiments. This idea has been further
explored qualitatively in a phenomenological spin-fermion model ~\cite
{aba}, in which the resonance mode is identified as the propagating
collective spin excitations and the scattering of electrons by these spin
modes gives rise to the anomalous spectral lineshape. Using the slave-boson
theory for the $t-t^{\prime }-J$ model, Brinckmann and Lee~\cite{bri} have
investigated the spin resonance and its evolution with doping.

Recent ARPES study reveals the specular dependences of the peak, dip and
hump energies with doping~\cite{cam}.  Furthermore, a small $\cos
(6\theta )$ ($\theta$ is the Fermi surface angle) deviation from the pure 
$d$-wave structure is observed \cite{Mesot}.  These results and their correlation
with the mode energy inferred from neutron data are the essential ingredient
for a consistent picture of the ARPES lineshape and the neutron data. In
this paper, we examine these issues and show that they can be quantitatively
reproduced based on the slave-boson approach to the 2D $t-t^{\prime }-J$
model.

The model reads, 
\begin{equation}
H=-\sum_{<ij>,\sigma }tc_{i\sigma }^{+}c_{j\sigma }-\sum_{<ij>^{\prime
},\sigma }t^{\prime }c_{i\sigma }^{+}c_{j\sigma }+{\frac{J}{2}}\sum_{<ij>}
{\bf S}_{i}\cdot {\bf S}_{j},
\end{equation}
where $<ij>$ denotes the nearest-neighbor (n.n.) bond and $
<ij>^{\prime }$ the next n.n. bond. In the slave-boson method, the physical
electron operators $c_{i\sigma }$ are expressed by slave bosons $b_{i}$
carrying the charge and fermions $f_{i\sigma }$ representing the spin; $
c_{i\sigma }=b_{i}^{+}f_{i\sigma }$. We consider the $d$-wave SC state with
the order parameters $\Delta _{ij}=<f_{i\uparrow }f_{j\downarrow
}-f_{i\downarrow }f_{j\downarrow }>$ and $\chi _{ij}=\sum_{\sigma
}<f_{i\sigma }^{+}f_{j\sigma }>$, in which bosons condense $b_{i}\rightarrow
<b_{i}>=\sqrt{\delta }$ ($\delta $ is the hole concentration)~\cite{ubb}.
Then, the mean-field Hamiltonian of Eq.(1) is, 
\begin{equation}
H_{m}=\sum_{k\sigma }\epsilon _{k}f_{k\sigma }^{+}f_{k\sigma }-\sum_{k}\Delta
_{k}(f_{k\uparrow }^{+}f_{-k\downarrow }+c.c)+2NJ^{\prime }(\chi
_{0}^{2}+\Delta _{0}^{2}),
\end{equation}
where the dispersion for fermions $\epsilon _{k}=-2(\delta t+J^{\prime }\chi
_{0})[\cos (k_{x})+\cos (k_{y})]-4\delta t^{\prime }\cos (k_{x})\cos
(k_{y})-\mu $, and the gap $\Delta (k)=2J^{\prime }\Delta _{0}[\cos
(k_{x})-\cos (k_{y})]$, with $J^{\prime }=3J/8$. The mean-field parameters $
\chi _{0}$, $\Delta _{0}$ and the chemical potential $\mu $ for different
doping $\delta $ are obtained from a self-consistent calculation~\cite{ubb}.

In the above mean-field calculation, we have set $<{\bf S}_{i}>$
=0~\cite{ubb}. In order to consider the response to external magnetic and
electric fields, we include the antiferromagnetic (AF) fluctuation by
writing $H=H_{m}+H^{\prime }$, where $H^{\prime }=H-H_{m}$, and
treating $H^{\prime }$ as a perturbation to $H_{m}$. In the
Hartree-Fock approximation, this reproduces the mean-field results~\cite{schriffer}. 
To investigate the AF fluctuations, we consider the scattering of
electrons off spin fluctuations in $H^{\prime }$. As a first step, we
calculate the spin susceptibility in the random-phase approximation (RPA) as
shown in Fig.1(a),
\begin{equation}
\chi ({\bf q},\omega )=\chi _{0}({\bf q},\omega )/[1+\alpha J({\bf q})\chi
_{0}({\bf q},\omega )].
\end{equation}
Here $J({\bf q})=J(\cos q_{x}+\cos q_{y})$ and $\chi _{0}({\bf q
},\omega )$ is the unperturbed spin susceptibility which is calculated
from the fermionic bubbles representing particle-hole excitations.
Following Ref.~\cite{bri}, we choose $\alpha =0.34$ in order to set the AF
instability at $\delta =0.02$, which is the experimental observed value.
This is the only adjusted parameter throughout the paper. The fermionic
self-energy is obtained from the lowest-order contribution of the
scatterings of fermions off spin fluctuations. In the SC state, there are
two different self-energies $\Sigma _{s}$ and $\Sigma _{w}$ as shown
in Fig.1(b) and (c), which renormalize the fermionic dispersion and the SC
gap, respectively. In the previous study~\cite{aba}, only $\Sigma _{s}$
is included in their calculations. We will show that the inclusion of $
\Sigma _{w}$ lead the otherwise flat doping dependence of the ARPES peak to
be consistent with experiments. The fermionic Green's function is calculated
by $G_{f}({\bf k},\omega )=[G_{f0}^{-1}({\bf k},\omega )+(\Delta _{k}+\Sigma
_{w})^{2}G_{f0}^{-1}(-{\bf k},-\omega )]^{-1}$ with $G_{f0}({\bf k},\omega
)=[i\omega -\epsilon _{k}-\Sigma _{s}({\bf k},\omega )]^{-1}$. In the SC
state, bosons condense and the physical electron Green's function can
be approximated by $G({\bf k},\omega )\approx \delta G_{f}({\bf q},\omega )$
. Then, the spectral function of electrons is calculated from $A({\bf k}
,\omega )=-(1/\pi ){\rm Im}G({\bf k},\omega )$. Numerical calculations are
performed at low temperature $T=0.005J$, with $t=2J,t^{\prime }=-0.45t$, and 
$J=0.13$eV.

We first analyze the imaginary part of the spin susceptibilities at 
${\bf Q}=(\pi ,\pi )$. It develops sharp resonance peaks for various
doping densities. This result has been reported in Ref.~\cite{bri}. In the
framework of the $d$-wave BCS theory, the origin of these peaks has
been discussed~\cite{bri,blu,li}. Essentially the peak arises from a
collective spin excitation mode corresponding to $1+\alpha J({\bf Q}){\rm Re
}\chi _{0}({\bf Q},\omega )=0$ and negligibly small Im$\chi _{0}({\bf
Q},\omega )$. It is due to a step-like rise of Im$\chi _{0}$ at
its gap edge and then a logarithmic singularity in Re$\chi _{0}$ via
the Kramers-Kroenig relation. This singularity shifts downward the
collective mode energy and leads it to situate in the spin gap, so no
damping is expected for the mode. The step-like rise in Im$\chi _{0}$
arises from the flat band(extended van Hove singularity) which is observed
near $(\pi ,0)$~\cite{dess} and the property that $\Delta _{{\bf k+Q}
}=-\Delta _{{\bf k}}$ for transition momentum ${\bf Q}$ due to
the $d$-wave gap symmetry~\cite{li}.

The lineshape $A({\bf q},\omega )f(\omega )$ ($f(\omega )$ is the
Fermi distribution function) of electrons coming from the scatterings by
spin fluctuations for doping $\delta =0.12$ are shown in Fig.2. The
results are plotted for several wavevectors $k=(\pi ,0),(\pi ,0.15\pi )$,
..., down to $(\pi ,0.5\pi )$. Clear peak/dip/hump structures are present at
and near $(\pi ,0)$. In order to understand the origin of the peak/dip/hump
structure, we plot the self-energy $\Sigma _{s}$ [$\Sigma _{w}$ is
qualitatively similar to $\Sigma _{s}$] at the Fermi wavevector $k_{F}$ for
doping $\delta =0.12$ in the inset of Fig.2. The solid line denotes
its real part $\Sigma _{s}^{^{\prime }}$, while the dashed line its
imaginary part $\Sigma _{s}^{^{\prime \prime }}$. The corresponding
lineshape at $k_{F}$ is expressed by the line with open squares in the main
panel. Due to the coupling to the spin resonance mode, the whole structure
of the self-energy $\Sigma _{s}$ is very similar to that of the unperturbed
spin susceptibility $\chi _{0}$ (for a comparison, see Fig.2 in Ref.~\cite
{li}). When frequency $|\omega |$ is below about 0.5$J$, $\Sigma
_{s}^{^{\prime \prime }}$ is equal to zero. Above 0.5$J$, a step-like rise
can be seen and is followed by a decrease. Consequently, $\Sigma
_{s}^{^{\prime }}$ has a peak at about the center of the step-like rise of $
\Sigma _{s}^{^{\prime \prime }}$. The quasiparticle energy is given by the
pole of the Green's function, which is the solution $\omega $ of the
equation $P({k_{F}},\omega )={\rm Re}[(\omega -\Sigma _{s}(k_{F},\omega))^{2}-
(\Delta_{k_{F}}+\Sigma _{w}({k_{F}},\omega ))^{2}]=0$. We also show
$P(k_{F},\omega )$ (dotted line) in the inset of Fig.2. The lowest
binding-energy solution of the pole equation $P(k_{F},\omega )=0$ is $\omega
=-0.42J$. Meanwhile, the damping of this mode which is proportional to $
\Sigma _{s}^{^{\prime \prime }}$ and $\Sigma _{w}^{^{\prime \prime }}$
approaches to zero. Therefore, it gives rise to a quasiparticle mode which
is denoted by the sharp low binding-energy peak in the lineshape shown in 
Fig.2. As $|\omega |$ increases further, the pole equation does not
satisfy anymore. Near the end of the step-like rise in $\Sigma
_{s}^{^{\prime \prime }}$, $P(k_{F},\omega )$ reaches its local maximum,
meanwhile the imaginary part of the self-energy is also near its maximum,
therefore a dip appears. It indicates that the dip is caused by the
step-like rise in the imaginary part of the self energy. After the dip, $
P(k_{F},\omega )$ decreases with $|\omega |$ and near $\omega =-1.0J$, it
reaches a local minimum. This leads to the broad higher binding-energy hump.

We now discuss the dispersion of the hump/dip/peak structure. At the
momentum range below and slightly above the Fermi wavevector $(\pi ,0.15\pi
) $, the low binding-energy peak nearly does not disperse with $k$. Above $
(\pi ,0.3\pi )$, the peak starts to move to higher binding-energy but cannot
move further than that dip seen around $k_F$. Then, it disappears gradually
with the further increasing of $k$, due to the unavailable quasiparticle states
above the Fermi surface at low temperatures. As a result, the dip also
disappears gradually. These behaviors were observed in experiments~\cite{cam}. 
The position of the peak determines the normalized gap size due to spin fluctuations. 
In Fig.3, we show a typical dependence of the renormalized gap
magnitude on the Fermi surface angle $\theta $, which is the angle between
$k_F$ and $k_x$. In general, the next order deviation from the d-wave starts 
from $\cos (6\theta )$ and can be described by the form $\Delta (\theta )=
\Delta _1\cos (2\theta )+\Delta_2\cos (6\theta )$. However, our results indicate
 that the best fit is $\Delta (\theta )=\Delta _{\max }[B\cos (2\theta )
+(1-B)\cos (6\theta )]$ with $B$ being around 0.88-0.94. This particular form was 
also used in Ref.\cite{Mesot} to fit their ARPES data with roughly the same range
of $B$.

In Fig.4(a), we show the doping dependence of the positions of the hump and
the peak, and compare our results with the experiment~\cite{cam}. One can
see that our result for the peak is in reasonably good agreement with
the experiment. We note that, if we just include the self-energy $\Sigma _{s}
$ in our calculation as was done in Ref.~\cite{aba}, a doping
independence for the peak positions is obtained. Hence the renormalization
of the gap by including $\Sigma _{w}$ is important. The magnitude of the
position of the hump and its trend with the doping variation are also
consistent with the experiment. However, the slope of the curve of hump
versus hole density is flatter than the experimental result.  
Considering that all parameters in our calculations are chosen
according to the well-known values and the only adjusted parameter $\alpha $
is fixed by the experimental observation on AF instability, these results
are quite satisfactory. The dependence of the ratio between the hump
and peak energy at $(\pi ,0)$ on doping concentrations is shown in Fig.4(b).
A flat variation for a wide doping level is seen. This result again agrees
with the experiment~\cite{cam}.

An important quantity addressed in the ARPES experiments is the peak-dip
separation, which is shown to be close to the mode energy of the neutron
peak~\cite{cam}. From a comparison of  Fig.2 in Ref.\cite{bri} and
Fig.3, one can see that the peak-dip separation in our calculations is lower
than the neutron peak. However, we would like to point out that this
separation is sensitive to the experimental energy resolution simulated to
be $\Gamma =0.02J$ here. When $\Gamma $ increases, the self-energy $\Sigma
_{s}^{^{\prime \prime }}$ decreases and the quasiparticle peak broads,
consequently the peak-dip separation increases. However the position
of the neutron scattering peak does not change for different $\Gamma $, it
just becomes broad. So, this comparison is also sensitive to $\Gamma $. But,
as noted above, the dip stems from the step-like edge in $\Sigma
_{s}^{^{\prime \prime }}$ which is in turn caused by the coupling to the
collective spin mode. On the other hand, this spin resonance mode also
arises from the step-like edge in Im$\chi _{0}$ which is due to the extended
van Hove singularity near $(\pi ,0)$ and the $d$-wave symmetry. Therefore,
an intimate relation between them is suggested. Thus, our result seems
to provide a consistent picture for the spin resonance peak and the
hump/dip/peak structure based on the spin excitations in a $d$-wave
superconductor.

Recently, P.J.White {\it et al.}~\cite{white} have investigated the effect
of nonmagnetic Zn impurities on the lineshape of $
Bi_{2}Sr_{2}Ca(Cu_{1-x}Zn_{x})O_{8+\delta}$ by ARPES experiment. They found
that the dip is diminished with Zn doping. According to the previous studies
by one of the authors (Li)~\cite{li}, Zn doping will wash out the van Hove
singularity and cause the decays of quasiparticle states. So, the
enhancement in Re$\chi_{0}$ and the step-like rise in Im$\chi_{0}$ will be
suppressed. As a result, no clear resonance neutron peak appears at certain
Zn doping concentration~\cite{li}. Because the dip is suggested to come from
the coupling to the spin resonance mode here, the disappearance of the dip
upon Zn doping may be naturally explained in the present framework.

It is quite encouraging that our results fits various kinds of ARPES and
neutron scattering data with correct trend and reasonable magnitude. There
is only one adjustable parameter which is fixed by the experimentally
observed AF instability. Therefore, our investigation represents a natural
extension to the work of Brinckmann and Lee~\cite{bri}, which addressed the
doping dependence of the resonance peak in neutron scatterings. In addition,
in comparison to the qualitative study by Abanov and Chubukov~\cite{aba},
our study further furnishes a quantitative basis.

In summary, based on the slave-boson approach to the $t-t^{\prime }-J$
model, we show that the anomalous peak/dip/hump structure observed in the
ARPES experiments arises from the coupling of quasiparticle to the
collective spin excitations which gives rise to the resonance peak in the
neutron scattering experiments. Our investigation seems to give a
consistent explanation for the resonance neutron peak and the ARPES spectra
based on the spin excitations in a $d$-wave superconductor.

We acknowledge the support from NSC of Taiwan under Grant
Nos.88-2112-M-001-004 and 89-2112-M-007-024. JXL was support in part
by the National Nature Science Foundation of China. He would also like to
thank Institute of Physics, Academia Sinica (Taiwan) for support during the
initial stages of this work and C.D.Gong for helpful discussion.

\vspace{1cm}

\section*{FIGURE CAPTIONS}

\vspace{0.3cm} Fig.1 Feynman diagrams for: (a) the RPA approximation to the
spin susceptibility coming from particle-hole excitations. (b) and (c) the
lowest-order contribution to the self-energy from fermion-spin excitation
scatterings.

\vspace{0.3cm} Fig.2 Photoemission spectra $A({\bf k},\omega )f(\omega
)$ at $k=(\pi ,0),(\pi ,0.15\pi ),(\pi ,0.225\pi ),(\pi ,0.3\pi ),\newline
(\pi ,0.4\pi ),(\pi ,0.5\pi )$( from the up to down lines). Inset shows the
frequency dependence of the fermion's self-energy $\Sigma_{s}$ at 
$k_{F}=(\pi ,0.15\pi )$ (Fermi wavevector). The solid line corresponds to 
its real part and the dashed line to its imaginary part. Also shown in the 
inset is $P(k_{F},\omega )$(see text) which is denoted by the dotted line.

\vspace{0.3cm} Fig.3 The renormalized gap size versus the Fermi surface angle 
$\theta $ for $\delta =0.08$. The open circles are our results, while the solid 
line is the fit using the gap function $\Delta (\theta )=\Delta _{\max }
[B\cos (2\theta )+(1-B)\cos (6\theta )]$ with $B=0.94$. 

\vspace{0.3cm} Fig.4 Doping dependences of the energy scales for the hump
and the peak at $(\pi,0)$. (a) Doping dependences of the hump and peak
positions. The open squares and the solid triangles are experimental data
for the hump and peak positions from Ref.~\cite{cam}, respectively. (b)
Doping dependence of the ratio of the hump position to the peak position.

\end{document}